\documentclass[
  journal=largetwo,
  manuscript=article-type,
  year=2020,
  volume=37,
]{cup-journal}

\usepackage{amsmath}
\usepackage[nopatch]{microtype}
\usepackage{booktabs}

\usepackage{multirow}
\usepackage{threeparttable}
\DeclareUnicodeCharacter{2212}{-}

\def\kms{\>{\rm ~km}\,{\rm s}^{-1}}

\def\Msun{\>{\rm M_{\odot}}}
\newcommand{\comment}[1]{}

\comment{
\definecolor{ref@color}{RGB}{0,76,153}
\usepackage{hyperref}
\hypersetup{colorlinks=true,linkcolor=ref@color,citecolor=ref@color,urlcolor=ref@color}
}

\title{Statistical analysis of kicked black holes from TNG300 simulation}

\author{M. Smole}
\affiliation{Astronomical Observatory, Volgina 7, 11060 Belgrade, Serbia}
\email[M. Smole]{msmole@aob.rs}

\author{M. Micic}
\affiliation{Astronomical Observatory, Volgina 7, 11060 Belgrade, Serbia}


\begin{document}

\begin{abstract}

Asymmetric emission of gravitational waves during mergers of black holes (BHs) produces
a recoil kick, which can set a newly formed BH on a bound orbit around the center of its host galaxy, 
or even completely eject it. 
To study this population of recoiling BHs  we extract properties of galaxies with merging BHs from
Illustris TNG300 simulation
and then employ both analytical and numerical techniques to model 
unresolved process of BH recoil.

This comparative analysis between analytical and numerical models shows that,
on cosmological scales, numerically modeled recoiling BHs have a higher escape probability 
and predict a greater number of offset active galactic nuclei (AGN).
BH escaped probability $>40~ \%$ is expected in 25 $\%$ of merger remnants in numerical models, 
compared to 8$\%$ in analytical models.
At the same time, the predicted number of offset AGN at separations 
$>5$ kpc changes from 58 $\%$ for numerical models to 3 $\%$ for analytical models.
Since BH ejections in major merger remnants occur in non-virialized systems,
static analytical models cannot provide an accurate description. 
Thus we argue that numerical models should be used to estimate the expected number density of 
escaped BHs and offset AGN.

\end{abstract}

\section{Introduction}

The concept of hierarchical growth of structures suggests that galaxy mergers 
play an important role in their evolution. 
Alongside building their masses, galaxy mergers can reshape their structure, 
enhance star formation and potentially trigger active galactic nuclei (AGN). 
Both theory and observations reveal the potential of major galaxy mergers 
to fuel enough gas to the central galaxy regions and increase the central black hole (BH)
accretion rate (e.g. \cite{cox}; \cite{lambas}; \cite{patton}; \cite{Weston}; \cite{Ellison}; \cite{montero}; \cite{Byrne}).

When two galaxies merge, the central black holes (BHs) gradually lose energy and move toward the center of the merged galaxy due to dynamical friction. This results in the formation of a binary BH, which can continue to harden through interactions with stars and gas.
The formation of a close BH binary system is significantly affected by the morphology 
and mass ratio of progenitor galaxies. Major mergers of gas-rich galaxies
are expected to result in effective hard BH binary formation
(\cite{Escala1, Escala2}; \cite{Mayer}; \cite{roskar};
\cite{delValle}; \cite{Goicovic}; \cite{Khan_2016}),
while some minor mergers do not necessarily lead to central BHs merger
(\cite{Callegari_2009, Callegari_2011}; \cite{Tremmel}).

In the final stages of BH merger, the emission of gravitational waves will effectively remove angular momentum and energy from the binary system, leading to a rapid collision \citep{Begelman}.
Those gravitational waves could be
directly detected with the upcoming space-based interferometer LISA (Laser Interferometer Space Antenna, \cite{Amaro}), whose focus will be on the mergers of BHs in the mass range
$10^{4}-10^{7} \Msun$.
Any asymmetry in the binary system, such as the presence of BHs with unequal masses or spins, can cause the asymmetric emission of gravitational radiation and BH recoil.
The emission of gravitational waves has a preferred direction due to a non-zero net linear momentum, and this causes the center of mass of the binary to move in the opposite direction \citep{Redmount}.
During this process, the newly formed BH receives a recoil whose amplitude depends on
the mass ratio of the merging BH,  the magnitude and orientation of their spins with respect to the binary's orbital plane, and the eccentricity of the orbit. Mergers of
fast rotating BHs with spin vectors in the orbital plane and orientated in opposite directions
result with super-kicks as high as $~4000\kms$, while efficient spin alignment during 
BH merger reduces kick velocities (\cite{gonz}; \cite{camp}; \cite{lousto}).

Depending on the amplitude of the kick velocity, a recoiling BH could be either placed on a bound orbit around the galaxy center or be completely ejected from its host.
Bound BHs can spend up to several Gyr oscillating with low amplitude around the galaxy center
(e.g., \cite{Gualandris}; \cite{Komossa}), 
while this time can be significantly reduced in gas-rich systems
(e.g., \cite{Blecha08}; \cite{Guedes}; \cite{Sijacki}).
In the case that merged BH receives a recoil velocity higher than the escape velocity from 
the host galaxy, it will be completely ejected.
Those wandering BHs are less likely to merge with other BHs, thus gravitational
wave recoil could have a negative influence on merger-driven BH growth (e.g., 
\cite{Haiman}; \cite{Merritt}; \cite{Volonteri}).
If active at the time of the merger, the accretion disk
will be ejected along with a recoiling BH, during which 
time it can be observed as spatially or kinematically offset AGN (e.g. \cite{Madau};
\cite{Loeb}; \cite{Blecha08}; \cite{Blecha11}).
There is a large number of observed offset AGN candidates in the literature
(e.g. \cite{Komossa1}; \cite{Robinson}; \cite{Jonker}; \cite{Civano10,Civano12};
\cite{Eracleous}; \cite{Koss}; \cite{Markakis}; \cite{Chiaberge} ;\cite{Kalfountzou}).
More recently, \cite{Stemo21} composed a catalog of 204 offset and dual AGN, using 
the Advanced Camera for Surveys Active Galactic Nuclei Catalog \citep{Stemo20}
of AGNs observed by Hubble Space Telescope.
However, none of these cases have been confirmed since alternative
explanations, such as a binary BH system with only one active BH, cannot be ruled out.

Even though cosmological simulations provide a powerful tool 
to investigate the population of merging galaxies,
the actual process of a BH merger cannot be simulated
due to the resolution limits. Instead, the results of cosmological
simulations are often combined with analytical 
models for unresolved processes. 
\citet{Blecha} investigated trajectories of recoiling
BHs using Illustris cosmological simulations (Vogelsberger et al. 2014a,b ; \cite{illu3};
\cite{illu4}).
By extracting properties of
merging BHs and progenitor galaxies directly from the
simulation, authors constructed analytical models of merger
remnant galaxies to estimate escape velocities, BH kick 
velocities and offset AGN distribution.

On the other hand, higher mass resolution achieved by simulations of 
isolated galaxies can provide more detailed insights into processes 
within the merger remnant galaxy.
\citet{smole19} showed that numerical models
provide a more realistic representation of merging galaxies compared to analytical models. 
Mass redistribution during interaction
can lead to a 25 $\%$ lower escape
velocities in numerical models of major merger remnant galaxies 
(for more details we refer to Methods).

In this work, we combine the results of a cosmological simulation on one side and 
isolated galaxy simulations on the other side. 
We apply the similar analysis 
described by \citet{Blecha} to state-of-the-art \textit{The Next Generation Illustris Simulations}\footnote{https://www.tng-project.org/} (Illustris TNG) project (\cite{Springel18}; \cite{Naiman}; \cite{Pillepich}; \cite{Nelson}; \cite{Marinacci}).
We employ the simulation with the largest volume, TNG300,
to extract properties of galaxies with merging BHs.
In addition to analytical models,
we used the results of numerical simulation of isolated galaxy mergers in order
to estimate escape velocities \citep{smole19}.
We investigate the difference between the predicted number 
of recoiling BHs in analytical and numerical models on cosmological scales
and estimate the probability for complete BH ejections during major galaxy mergers.

In Methods we introduce the employed 
techniques, together with data sets used in this work,
and we describe their implementation
in our methodology.
Our findings are presented and discussed in Results.
In Conclusions we summarise the main findings
of this work and draw conclusions.

\section{Methods}
\label{method}

\subsection{Illustris TNG300}

In order to calculate the statistics of recoiling BHs in  numerical and analytical models, 
we use publicly available data from Illustris TNG project (\cite{Springel18}; \cite{Naiman}; \cite{Pillepich}; \cite{Nelson}; \cite{Marinacci}).
Illustris TNG cosmological hydrodynamical simulations of galaxy formation
were performed using the \textsc{arepo} code (\cite{Springel10}).
The simulations were initialized at redshift $z = 127$ using  \citet{planck} cosmological parameters: matter density $\Omega_{\rm{m}}=0.3089$, baryon density $\Omega_{\rm{b}}=0.0486$, dark energy density $\Omega_{\Lambda}=0.6911$,  Hubble constant $H_{\rm{0}} = 0.6774$ $\rm{km}/\rm{s}/\rm{Mpc}$,
power spectrum normalization $\sigma_{\rm{8}}=0.8159$ and a primordial spectral index $n_{\rm{s}} = 0.9667$.
For the analysis employed in this work we choose the simulation with the largest cosmological box $(\sim 300~ \rm{Mpc})^{3}$, TNG300. Mass resolution for TNG300 simulation is  $5.9 \times 10^7 \Msun$ for DM particles and $1.1 \times 10^7 \Msun$ for baryonic component. 

Results of Illustris TNG300 simulation are stored in 100 snapshots, from redshift $z=20$ to $z=0$.
Snapshot files contain information about each particle in the simulation volume. 
In TNG300 simulation a BH particle
with a seed mass of $1.18\times10^6\Msun$ is placed in each galaxy more massive than $7.8 \times 10^{10}\Msun$, which do not already have a BH, and then allowed to grow via gas accretion and by mergers. BH accretion is modeled according to 
Eddington-limited Bondi accretion rate:

\begin{subequations}
\begin{equation}
 \dot{M}_\textrm{Bondi} = \frac{4 \pi G^2 M_\text{BH}^2 \rho}{c_s^3}, 
 \end{equation}

 \begin{equation}
 \dot{M}_\textrm{Edd} = \frac{4 \pi G M_\textrm{BH} m_p}{\epsilon_r \sigma_{\rm T}} c, 
 \end{equation}
 
 \begin{equation}
\dot{M} = \min\left( \dot{M}_\textrm{Bondi} , \dot{M}_\text{Edd}  \right),
\end{equation}
\end{subequations}
\noindent where $M_\text{BH}$ is BH mass, $G$ is the gravitational constant, $\rho$ the kernel-weighted ambient density around the BH, $c_s$ the kernel-weighted ambient sound speed including the magnetic signal propagation speed, $m_p$ the proton mass, $c$ the speed of light, $\epsilon_r = 0.2$ the BH radiative efficiency and $\sigma_{\rm T}$ the Thompson cross-section. We refer to \citet{Weinberger}
for more details about BH growth and the feedback processes included. 

An additional supplementary data catalog is available providing details of each BH-BH merger in the simulation,
otherwise not available from the snapshots alone. 
This file contains information about both merging BH ID numbers (unique throughout the simulation),
their masses immediately preceding the merger and snapshot during which, or immediately following, this merger event occurred.

The total number of BH merger events in Illustris TNG300 simulation is 590328. First, we exclude all multiple mergers,
that is all BHs that participate in more than one merger event per snapshot. 
In a simplified case during triple mergers,
the lowest mass BH will be ejected while the remaining BHs
will form a binary system. However, the dynamic 
of multiple BH systems requires complex treatment which is beyond the scope of this paper.
Even though this excludes 144048 merger pairs ($\sim~24\%$), our main goal is to 
obtain the comparative statistics between analytical and numerical predictions. 
Thus, we limit our analysis to a smaller sample of single BH mergers for which we have
well-studied analytical and numerical models.

The next step was to link all of the merging BHs from BH mergers supplementary catalog  with the associated subhaloes, i.e. the host galaxies. 
First, we identify the host galaxies of each BH participating in the
merger event at the snapshot preceding the merger, while BHs still occupy separate systems. 
Next, we follow backward the evolution of each host galaxy, from the snapshot preceding the BH merger to the snapshot when the galaxies first appeared in the simulation.
This step is necessary in order to accurately determine the mass ratio 
of the merging galaxies, since in the snapshot preceding the BH merger host galaxies have already exchanged 
mass as a consequence of galaxy interaction.
The mass ratio of the merging galaxies is calculated at the snapshot where the secondary, i.e. less massive galaxy, has
the maximum mass, thus prior mass loss due to galaxy merger. That snapshot is taken as the last snapshot before the galaxy interaction started. 
However, since a galaxy merger is a prolonged process, it is possible to imagine a scenario where
one of the progenitor galaxies undergoes 
a new merger between the last snapshot before galaxy interaction and the snapshot of the BH merger. 
That merger would change the total mass and other properties of the progenitor galaxy.
We limit our analysis to single mergers, thus we eliminate from our sample not only galaxies that undergo multiple mergers per snapshot but also galaxies that experience multiple mergers during galaxy interaction.

The additional selection criterion is imposed to ensure that each host galaxy is well resolved. Following \citet{Blecha} and \citet{Kelley} we require
that each progenitor galaxy, at the last snapshot before the galaxy
interaction, has $M_\textrm{DM}>10^{10}\Msun$ and $M_\textrm{star}>10^8\Msun$. In addition,
we also limit the upper mass of the host galaxy to $M_\textrm{DM}<8\times10^{13}\Msun$ 
to exclude galaxy clusters (\cite{Paul}).

As we will discuss in the following subsection, \citet{smole19} 
showed that escape velocities from minor merger remnants in analytical and numerical
models do not differ significantly.
In addition, since BH mass is expected to scale with the host galaxy mass, minor mergers would lead 
to the formation of low mass ratio BH binaries, and the expected kick velocity 
in such systems is not enough to eject the central BH (see the following subsections). 
Thus we limit our analysis to major merger remnants.
Our final sample contains 46031 major merger remnants or $\sim8~\%$ of the total number of all BH mergers in TNG300 simulation.

\subsection{BH escape velocities in analytical and numerical galaxy models}
\label{vesc}

\citet{smole19} investigated differences between recoiling BHs in analytical
and numerical potential using isolated N-body simulations performed with GADGET-2 code \citep{gadget}.
The authors constructed numerical models for a set of progenitor galaxies with different 
total masses and mass distributions within galaxy components and
simulated galaxy mergers with various galaxy mass ratios. 
In addition, the authors constructed analytical models of merger remnant galaxies 
with the same characteristics as their numerical counterparts. 
The main goal was to compare escape velocities from numerical and analytical merger remnants.
In numerical models, BH was represented as one massive particle placed at the merger remnant center
and its trajectory was  followed directly from the simulation.
In analytical models BH trajectory was
numerically integrated, using leapfrog integration.
Both analytical and numerical techniques result in the same value of escape velocity 
for isolated and virialized galaxies. However, 
galaxy mergers cause mass redistribution within merger remnants
which reduces escape velocities in numerical models. 
During galaxy mergers energy of individual particles is not conserved.
In the process called violent relaxation \citep{Lynden-Bell} weakly bound particles can escape the host
galaxy potential, which results in mass loss during mergers.
This process is not depicted in static analytical potential, making
the evolving numerical model a more realistic description of dynamical processes in
galaxies with merging BHs. 
As a consequence, \citet{smole19} showed that analytical models overestimate BH escape velocities.
BH escape velocities in numerical major merger remnant galaxies can be up to 25 $\%$ lower compared 
to those in analytical models.
However, this effect is predominantly limited to major mergers.
During minor mergers, the secondary galaxy  will experience significant mass
redistribution, but its contribution to the total potential 
is not sufficient to influence escape velocities, leading to the similar 
escape velocities in numerical and analytical minor merger remnants.

Here, we use the results of the above work to make comparative statistics
between recoiling BHs in TNG300 simulation, assuming analytical versus numerical models. 
We extract properties of merger remnant galaxies directly from the simulation.
For the given merger remnant mass, its central BH mass, and the mass
ratio of the merging galaxies, analytical and numerical escape velocities calculated by \citet{smole19} are
extrapolated and assigned to each galaxy from our TNG300 sample.

\subsection{BH kick velocities}
\label{vkick}

After assigning escape velocities to each major merger remnant galaxy, the next step is to calculate 
kick velocities for each newly formed BH, adopting the method described by \citet{micic11}.
Kick velocity depends on the merging BH mass ratio, their spin amplitude, and the alignment to the orbital angular momentum:
\begin{subequations}

\begin{equation}
V_{\mathrm{k}} = [(V_{m}+V_{\perp} \rm{cos}{\xi})^2+(V_{\perp} \sin{\xi})^2+(V_{\parallel})^2]^{1/2},  
\end{equation}
where
\begin{equation}
V_{m} = A \frac{q^2 \left(1-q\right)}{\left(1+q\right)^5} \left[ 1+
  B \frac{q}{\left(1+q\right)^2}\right], 
\end{equation}
\begin{equation}
V_{\perp} = H \frac{q^2}{\left(1+q\right)^5} \left( \alpha_2^\parallel
  - q \alpha_1^\parallel\right), 
\end{equation}
and
\begin{equation}
V_{\parallel} = K \cos\left(\Theta-\Theta_0\right) \frac{q^2}{\left(1+q\right)^5}
\left( \alpha_2^\perp - q \alpha_1^\perp \right).
\end{equation}
\end{subequations}

\noindent The fitting constants are A$=1.2 \times 10^{4} \kms$, B$=-0.93$, H$=(7.3 \pm 0.3) \times 10^{3} \kms$
and K$\cos\left(\Theta-\Theta_0\right) =(6.0 \pm 0.1) \times 10^{4} \; $; 
$q$ is the mass ratio of the merging BHs, $\alpha_{i=1,2}=\frac{S_{i}}{M_{i}}$ is the reduced 
spin parameter, $S_{i}$ is the spin angular momentum of BH, and the orientation of the merger is 
determined with angles $\Theta$ i $\xi$. 
Following \citet{micic11} we calculate kick velocities for two distinct spin distributions.
In both models, BH spin amplitudes are taken from
a uniform distribution, while BH spin orientation depends on a model choice. 
In the first model, BH spin orientations are also taken from
a uniform distribution ('random' model), while in the second model, BH spins
are always aligned with the orbital angular moment of the binary ('aligned' model), 
which results in lower kick velocities. 
The mass ratio of the merging BHs is taken directly from the BH merger supplementary catalog.
BH kick velocities are sampled from one of those
distributions using 10000 realizations for each merging BH pair from our sample.

\subsection{Escaped BHs versus Offset AGN statistics}

Next, we compare the sampled kick velocities
to the escape velocity from the merger remnant. Throughout this work we will 
refer to realizations with $v_{\textrm{k, i}}>v_{\textrm{esc}}$ ($i\in[1,...,10000]$)  as escaped BHs.
The escape probability of a host galaxy is defined as the fraction of escaped BHs out of 10000 realizations.
Recoiling BHs with $v_{\textrm{k, i}}<v_{\textrm{esc}}$ will not be ejected from the galaxy, but they can 
spend an extended period of time oscillating around the galaxy center. We refer to those as BHs on bound orbits.
We use the term offset AGN to refer to both escaped BHs and BHs on bound orbits, thus
all BHs with kick velocities large enough to
displace BH from the center of the galaxy.

In this work, we separately investigate the statistics of escaped BHs and offset AGN.
Our main goal is to calculate the comparative statistics between analytical and numerical predictions
on cosmological scales, using a sample of well-resolved major merger remnants
from TNG300 simulation, that reside outside of galaxy clusters.

\section{Results}
\label{results}

\subsection{Major merger remnants in TNG300}

\begin{figure}
	\includegraphics[width=0.9\columnwidth]{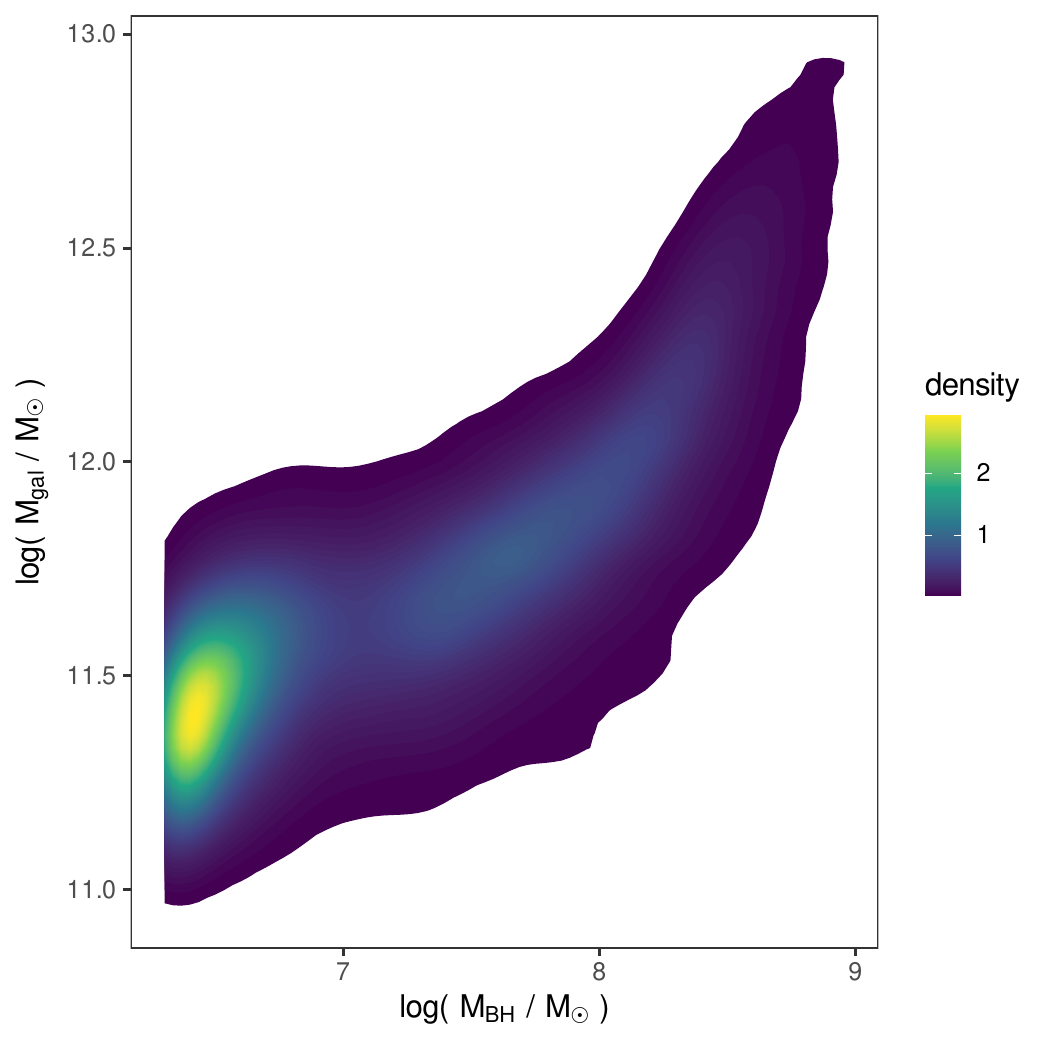}
    \caption{Total mass of progenitor galaxy as a function of the central BH mass. 
}
    \label{mgal-mbh}
\end{figure}

Fig.~\ref{mgal-mbh} shows the total mass of each progenitor galaxy as a function of the central 
BH mass. The color bar indicates the estimated 2D density of data points. Our sample is dominated by lower mass galaxies, $M_{\textrm{gal}}< 5\times10^{11}\Msun$, that host central BHs with masses  $M_{\textrm{BH}}<10^7\Msun$.

\begin{figure}
	\includegraphics[width=\columnwidth]{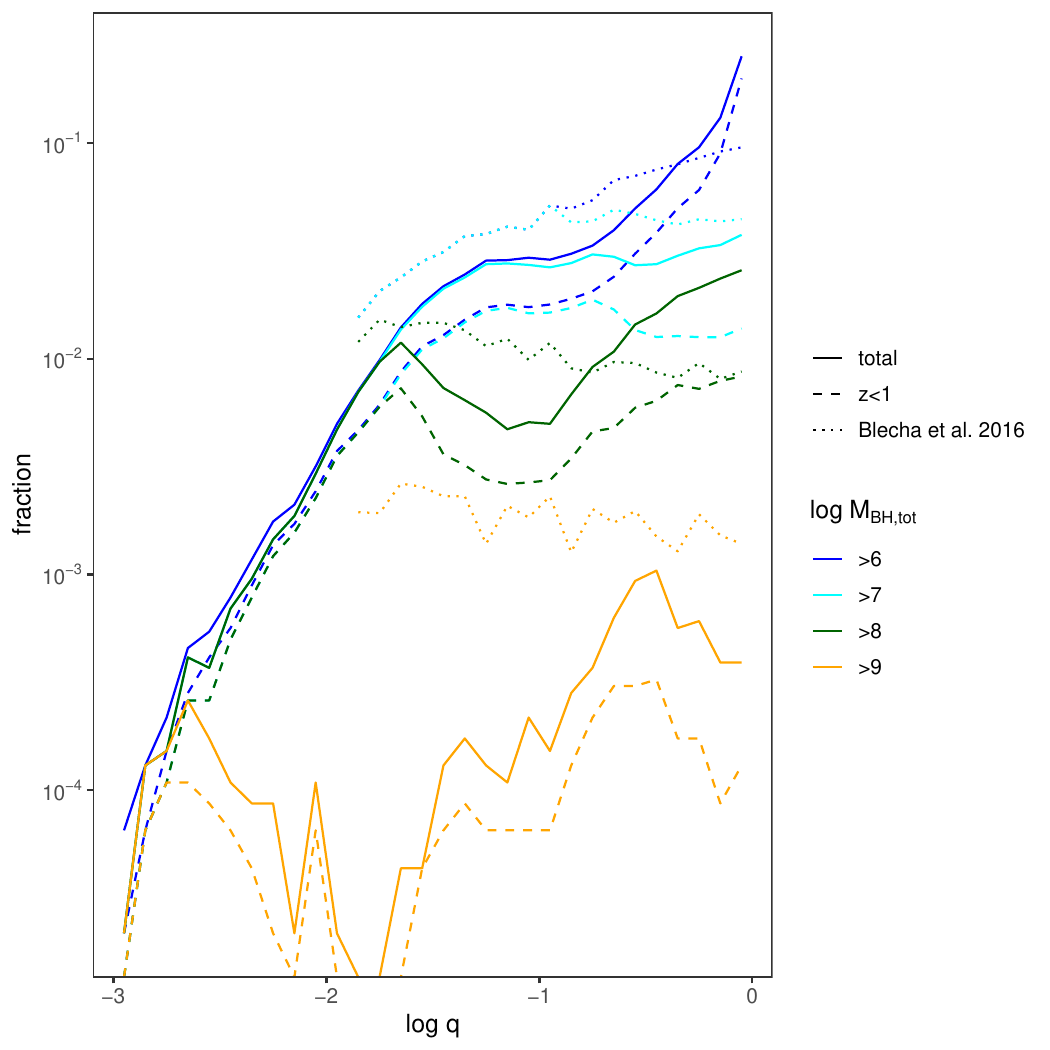}
    \caption{The distribution of merging BH mass ratio, for different total BH mass bins.
   Dashed lines represent the same distribution calculated for BH mergers at $z<1$. Dotted
   lines show the distribution in Illustris simulation, calculated by Blecha et al. 2016.
}
    \label{q_frac}
\end{figure}

Fig.~\ref{q_frac} shows the distribution of merging BH mass ratio, divided by the total 
BH mass, for all mergers (solid lines) and for mergers occurring at $z<1$ (dashed lines).
A lower fraction of high mass ratios at $z<1$ indicates that mergers of equal mass BHs 
were more common at high redshifts before multiple mergers and accretion led to 
BH growth. The distribution of all BHs (blue solid line), shows an increase 
toward equal mass mergers, while massive BHs show a flatter distribution.
Dotted lines show merging BH mass ratios in Illustris simulation, calculated by \cite{Blecha}. 
This distribution is flatter and with a higher fraction of low mass ratios.
Assuming that central BH mass scales with the host galaxy mass, choosing
only major galaxy mergers, as we did in our sample, would lead to higher BH mass ratios. 
A lower fraction of the most massive BHs reflects 
the imposed higher limit for total galaxy mass of $M_\textrm{DM}<8\times10^{13}\Msun$, so we 
exclude galaxy clusters. 

\begin{figure}
	\includegraphics[width=\columnwidth]{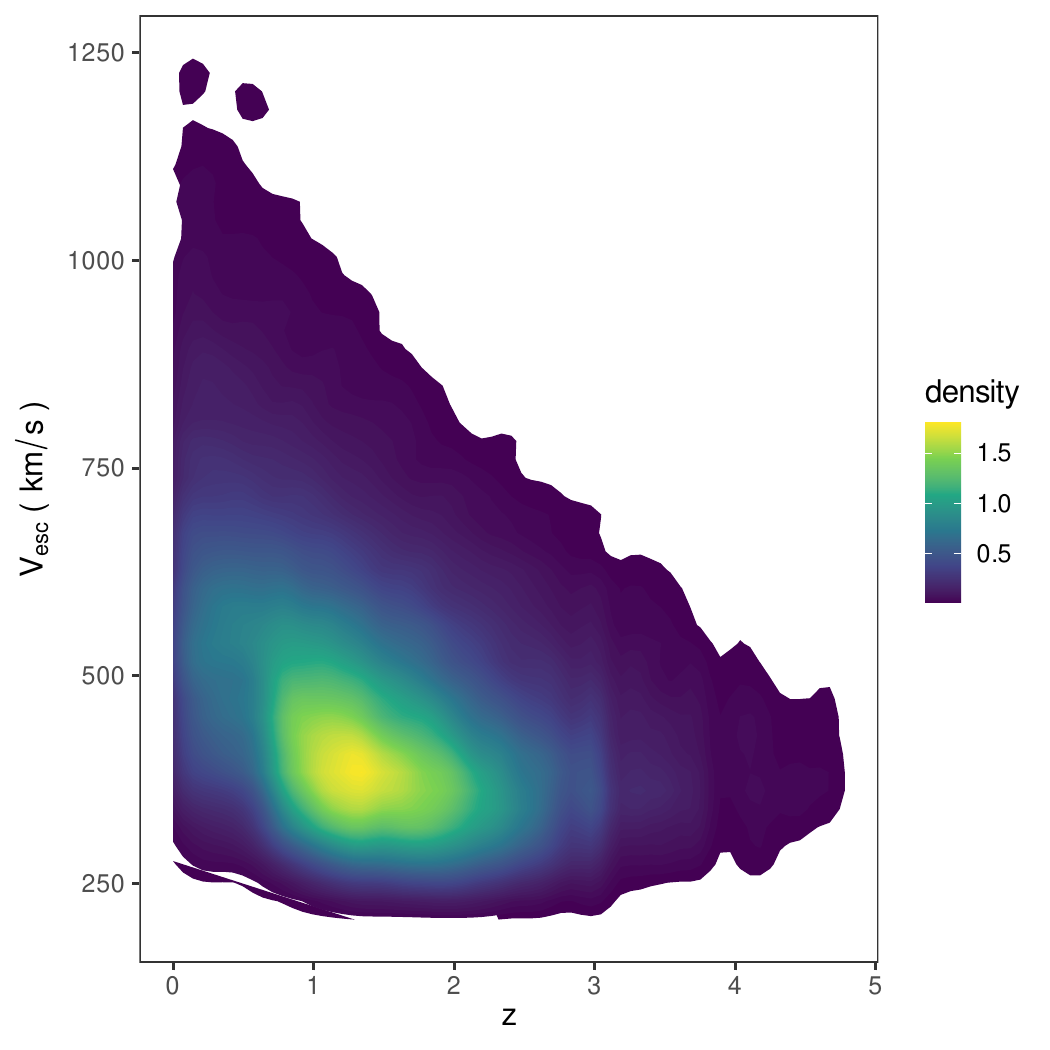}
    \caption{Numerically calculated escape velocity as a function of redshift.}
        \label{z_vesc}
\end{figure}

BH mergers at high redshifts are expected to produce more escaped BHs due to 
generally lower masses of merger remnant galaxies, and thus lower escape velocities. 
Fig.~\ref{z_vesc} shows escape velocities as a function of redshift, calculated using
numerical models. 
As predicted by the hierarchical growth of structures, the number of high-mass galaxies with high escape velocities decreases with redshift as galaxies build their masses and thus potential wells deep 
enough to retain a recoiling BH.

\begin{figure}
	\includegraphics[width=\columnwidth]{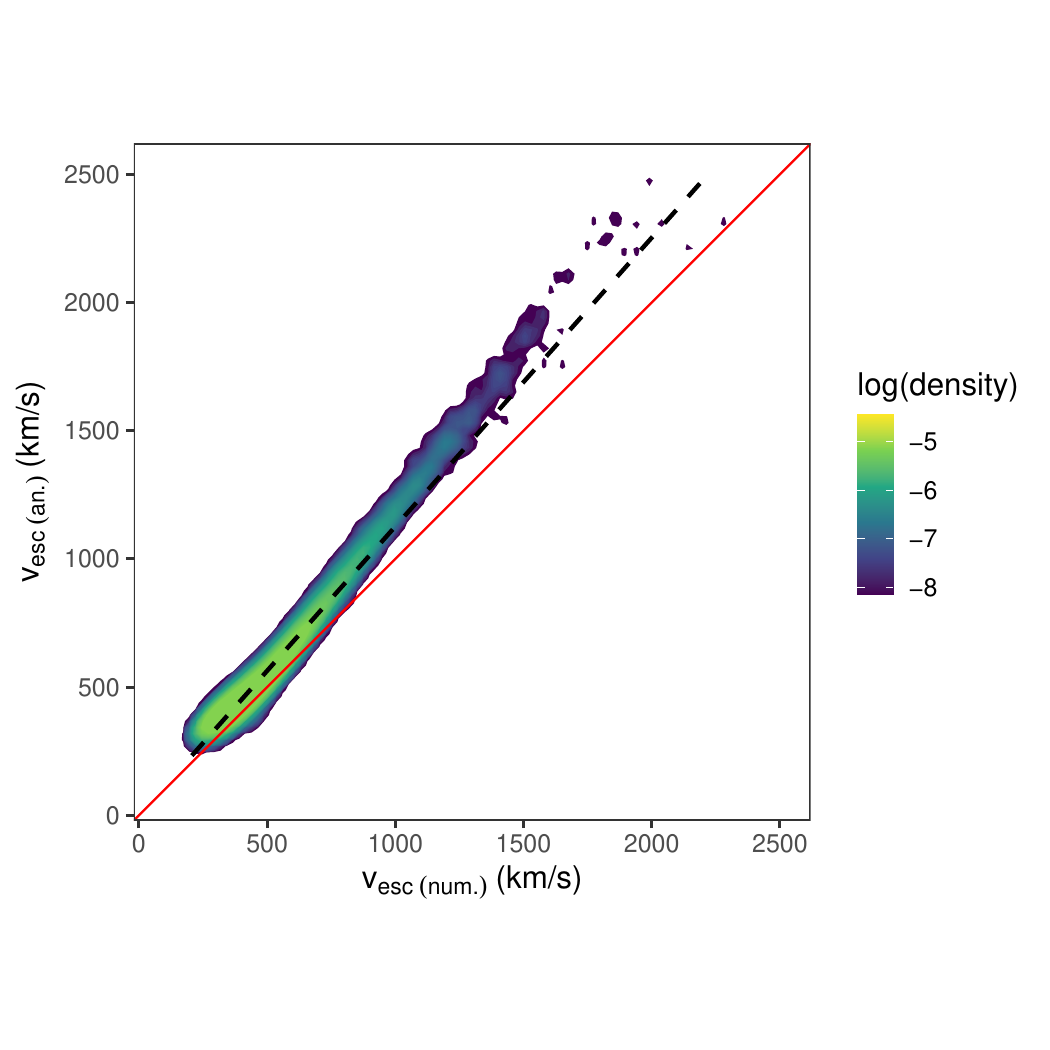}
    \caption{Escape velocities from galaxies in analytical models
as a function of their escape velocities in numerical models.  
The black dashed line represents a linear fit to our data and the solid red line denotes 
$v_\textrm{esc (an.)}=v_\textrm{esc (num.)}$.}
    \label{vesc_an_num}
\end{figure}

For the extracted galaxy mass, its central BH mass, and the mass
ratio of the merging galaxies we estimate the escape velocity for each galaxy in our sample using the results obtained 
by \citet{smole19}. 
Fig.~\ref{vesc_an_num} shows escape velocities from galaxies in analytical models
as a function of their escape velocities in numerical models. 
The color bar indicates the estimated 2D density of data points on a logarithmic scale. The black dashed line represents a 
linear fit to our data, while the solid red line denotes 
$v_\textrm{esc (an.)}=v_\textrm{esc (num.)}$.
The majority of galaxies from our sample occupy the region above the red line, thus on a statistical level
numerical models of major merger remnants predict lower escape velocities
compared to analytical models. 

\begin{figure}
	\includegraphics[width=\columnwidth]{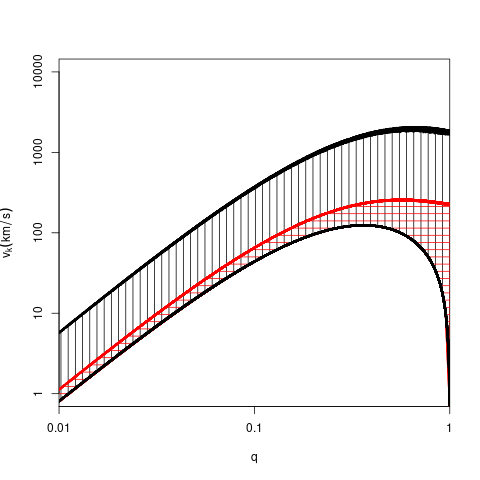}
    \caption{Distribution of kick velocities as a function of the mass ratio of merging BHs. Black vertical lines represent random spin model, while red horizontal lines denote to aligned spin model.}
    \label{vkick_plot}
\end{figure}

In this work, we adopted the method described by \citet{micic11} to calculate
kick velocities of merged BHs. 
Fig.~\ref{vkick_plot} shows the distribution of kick velocities from our data set, as a function of the merging BH mass ratio.
Values of kick velocities are the result of 10000 realizations per merger event, in which BH spins are sampled from random or from aligned distribution.
If BH spins prior BH merger are drawn from a uniform distribution (random model, black vertical lines)
the resulting kick velocities can be $> 2000~\textrm{km/s}$, which is enough to eject central BH even from the most massive galaxies in our sample.
On the other hand,  if BH spins are always aligned with the orbital angular moment of the binary
(aligned model, red horizontal lines), the highest possible kick velocities are $\sim 250~\textrm{km/s}$
and recoiling BH are expected to be rare. Both distributions yield the highest kick velocities
for mergers of BHs with comparable masses.

\subsection{Statistics of escaped BHs}

\begin{figure}
	\includegraphics[width=\columnwidth]{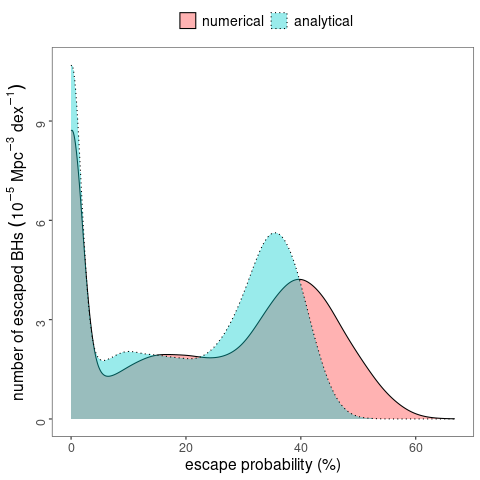}
    \caption{Probability function for a BH to escape for
random spin model of BH kick velocity distribution. Different colors
indicate numerical (solid line, pink) or analytical (dashed line, cyan) values for escape velocity calculation.
}
    \label{offset_random}
\end{figure}

Next, we calculate the probability that each BH merger pair from our sample produces
a kick large enough for complete BH removal. 
For each merger event, BH kick velocities are sampled from either random or aligned
spin distribution using 10000 realizations. 
Escape probability, or the probability for each merger event to completely eject BH, is
calculated by comparing each of the sampled kick velocities $v_{\textrm{k, i}}$ ($i\in[1,...,10000]$) 
to the escape velocity from the merger remnant. 
Realizations with $v_{\textrm{k, i}}>v_{\textrm{esc}}$  are referred to as escaped BHs.

Fig.~\ref{offset_random} shows BH escape probability
adopting random spin model for BH kick velocity distribution.
Different colors correspond to different models used to estimate the escape velocity from the merger remnant galaxy, numerical (solid line, pink) and analytical (dashed, cyan). 
If the BH spin parameters are taken from a uniform distribution, the maximal probability for 
escaped BH is
$\sim67~\%$ and $\sim55~\%$ for numerical and analytical models, respectively. Both models predict that the majority of merger events from our sample would not lead to BH removal, however, both distributions show a local maximum value 
around $\sim~40\%$ (numerical) and $\sim35~\%$ (analytical). 
Thus, the plot shows that the model with numerically estimated escape velocities results in a higher probability for escaped BHs compared to the analytical model.

Under the assumption that BH spins are always aligned with the orbital angular moment of the binary (aligned spin model), resulting kick velocities are lower (Fig~\ref{vkick_plot}). In fact, our sample 
does not contain any escaped BHs for the analytical model for escape velocity, while
numerical model predicts the highest probability of $\sim2~\%$ for a merger event to 
remove BH. Thus, if BH spin alignment during BH mergers is efficient, 
ejection of a newly formed BH from its host potential well is not excepted.

Further, we explore how the probability that a BH merger will produce an escaped BH evolves with time. 
Fig.~\ref{offset_random_z} shows BH probability to escape for different redshift bins. 
At redshifts $z>5$ escaped BH probability function peaks close to 40 $\%$, followed 
by a bimodal distribution at redshifts $1<z<4$, with the second peak at zero probability.
At low redshifts the function shows one peak and our sample is dominated 
by BH mergers that will not lead to BH removal.
Again, for each redshift bin escaped BH probability function is shifted toward higher values when numerical models are used.

\begin{figure}
	\includegraphics[width=\columnwidth]{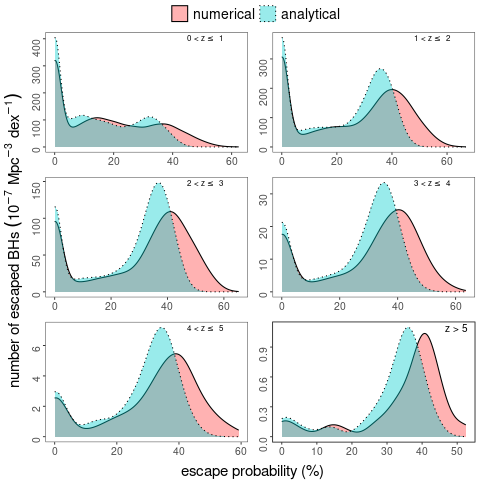}
    \caption{Same as Fig.~\ref{offset_random}, for different redshift bins.
}
    \label{offset_random_z}
\end{figure}

\comment{


            \begin{figure*}
   \resizebox{\hsize}{!}{
\includegraphics[width=1\textwidth]{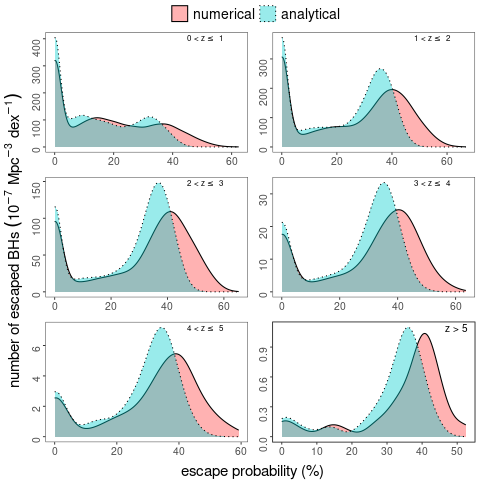}}
       \caption{Same as Fig.~\ref{offset_random}, for different redshift bins.
}
         \label{offset_random_z}
   \end{figure*}

}

Table~\ref{escaped_bhs} shows the total number of BH mergers that have escape probability $>20~\%$ and $>40~\%$, assuming
analytical versus numerical potential. Random spin model predicts that over $50 \%$ of BH mergers from our sample, 
produce a BH with escape probability $>20~\%$ for both numerical and analytical models. However, at higher 
escape probabilities the difference between our models becomes significant. 
Numerical models predict that $25 \%$ of BH mergers leave a remnant BH with escape probability $>40~\%$, compared 
to $8~\%$ in analytical models. Under the assumption that BH spins are aligned, escape probabilities $>2~\%$ are not expected.

Similarly, Table~\ref{escaped_bhs_z} shows the total number of BH mergers with 
escape probability $>20~\%$ and $>40~\%$, per redshift bin. The results presented here refer to random model of spin distribution. 
At each redshift bin, numerical models produce 
a higher fraction of BH mergers with escape probability exceeding $40~\%$. Moreover,
the gap between the fraction of escaped BHs predicted analytically versus numerically widens as redshift increases.
This is a consequence of generally more turbulent galaxy evolution at higher redshifts, with numerous mergers occurring before galaxies
acquire sufficient mass to retain the majority of recoiled BHs.
Numerical models better capture these processes in merging galaxies
and thus provide a more accurate estimate of the number of escaped BHs.
If an offset AGN is defined as a kicked BH with escape probability $>40~\%$, numerical 
models predict a substantially more significant number of BH ejected from the host galaxy centre, compared to analytical models. 
However, the gap between the two models becomes less pronounced when considering an escape probability $>20~\%$.

\begin{table}[hbt!]
\begin{threeparttable}
\caption{Total number of escaped BHs}
\label{escaped_bhs}
\begin{tabular}{|c|c|cc|}
\hline
\multirow{-1}{*}{spin } & \multirow{-1}{*}{escape} & \multicolumn{2}{c|}{total number of escaped BHs} \\ \cline{3-4} 
  model & probability & \multicolumn{1}{c|}{analytical} & numerical \\ \hline
 & \textgreater 20\% & \multicolumn{1}{c|}{24456 (53\%)} & 26510 (57\%) \\ \cline{2-4} 
\multirow{-2}{*}{random} & \textgreater 40\% & \multicolumn{1}{c|}{3741 (8\%)} & 11580 (25\%) \\ \hline
 & \textgreater 20\% & \multicolumn{1}{c|}{0} & 0 \\ \cline{2-4} 
\multirow{-2}{*}{aligned} & \textgreater 40\% & \multicolumn{1}{c|}{0} & 0 \\ \hline
\end{tabular}
\end{threeparttable}
\end{table}

\begin{table}[hbt!]
\begin{threeparttable}
\caption{Total number of escaped BHs per redshift bin}
\label{escaped_bhs_z}
\begin{tabular}{|c|c|cc|}
\hline
\multirow{-1}{*}{redshift} & \multirow{-1}{*}{\begin{tabular}[c]{@{}c@{}}escape \end{tabular}} & \multicolumn{2}{c|}{total number of escaped BHs} \\ \cline{3-4} 
 & probability & \multicolumn{1}{c|}{analytical} & numerical \\ \hline
$0 < z \le 1$ & \textgreater 20\% & 5064 (34\%) & 6106 (40\%) \\ \cline{2-4} 
 & \textgreater 40\% & 257 (2\%) & 1594 (11\%) \\ \hline
$1 < z \le 2$ & \textgreater 20\% & 11410 (58\%) & 12126 (62\%) \\ \cline{2-4} 
 & \textgreater 40\% & 1907 (10\%) & 5680 (29\%) \\ \hline
$2 < z \le 3$ & \textgreater 20\% & 6170 (69\%) & 6404 (72\%) \\ \cline{2-4} 
 & \textgreater 40\% & 1324 (15\%) & 3427 (39\%) \\ \hline
$3 < z \le 4$ & \textgreater 20\% & 1442 (70\%) & 1491 (72\%) \\ \cline{2-4} 
 & \textgreater 40\% & 214 (10\%) & 715 (35\%) \\ \hline
$4 < z \le 5$ & \textgreater 20\% & 308 (74\%) & 321 (77\%) \\ \cline{2-4} 
 & \textgreater 40\% & 32 (7\%) & 136 (33\%) \\ \hline
$z > 5$ & \textgreater 20\% & 39 (85\%) & 39 (85\%) \\ \cline{2-4} 
 & \textgreater 40\% & 4 (9\%) & 19 (41\%) \\ \hline
\end{tabular}
\end{threeparttable}
\end{table}

\begin{figure}
	\includegraphics[width=\columnwidth]{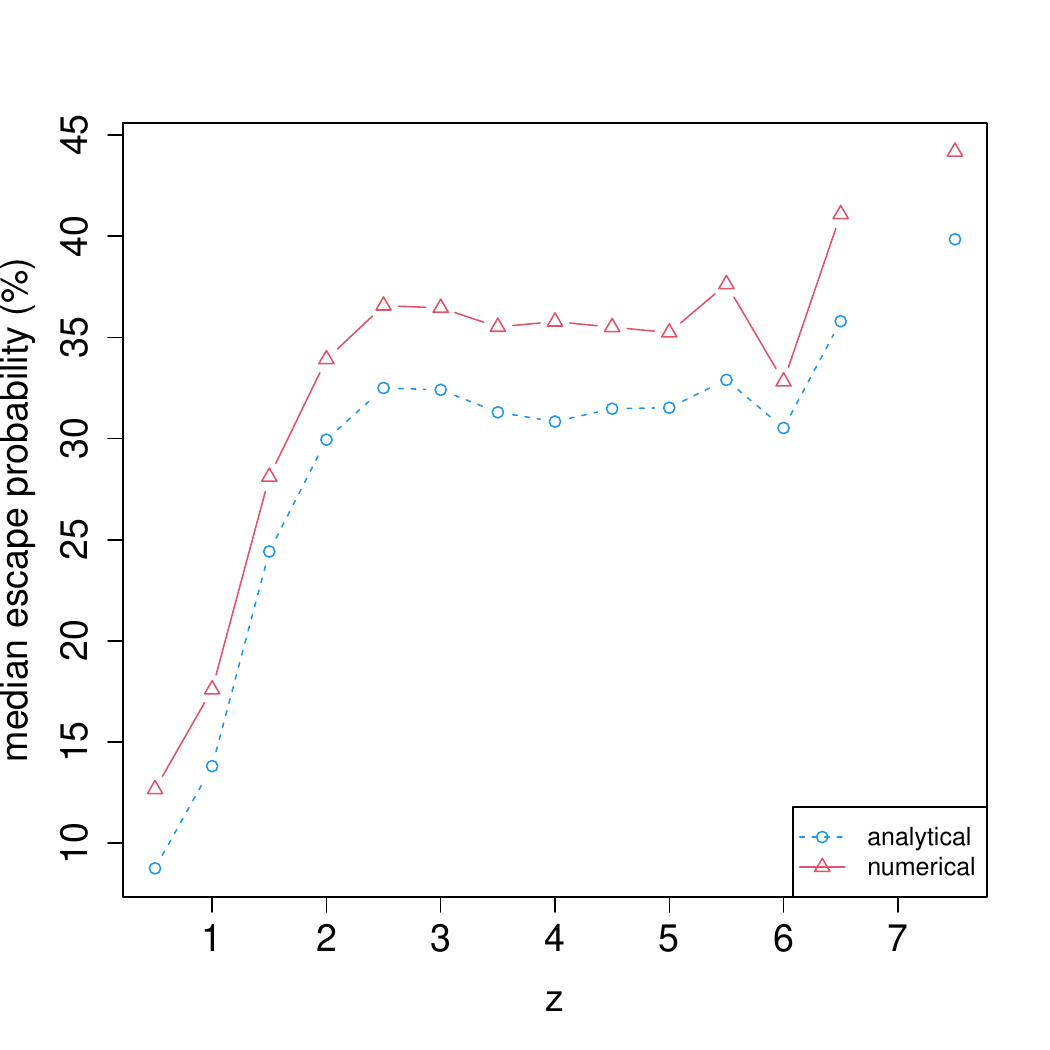}
    \caption{Median escape probability as a function of redshift.
Kick velocities are calculated assuming random spin distribution.}
    \label{median}
\end{figure}

Fig.~\ref{median} shows the median escape probability  as a function of redshift, for numerical (red) and analytical (blue) models. 
The median probability that a BH merger pair would produce an escaped BH increases with redshift.
At high redshifts a greater percentage of BH mergers are expected to result in kick velocities high enough to eject central BH.

\subsection{Statistics of offset AGN}

  \begin{figure*}
   \resizebox{\hsize}{!}
            {\includegraphics[width=.45\textwidth]{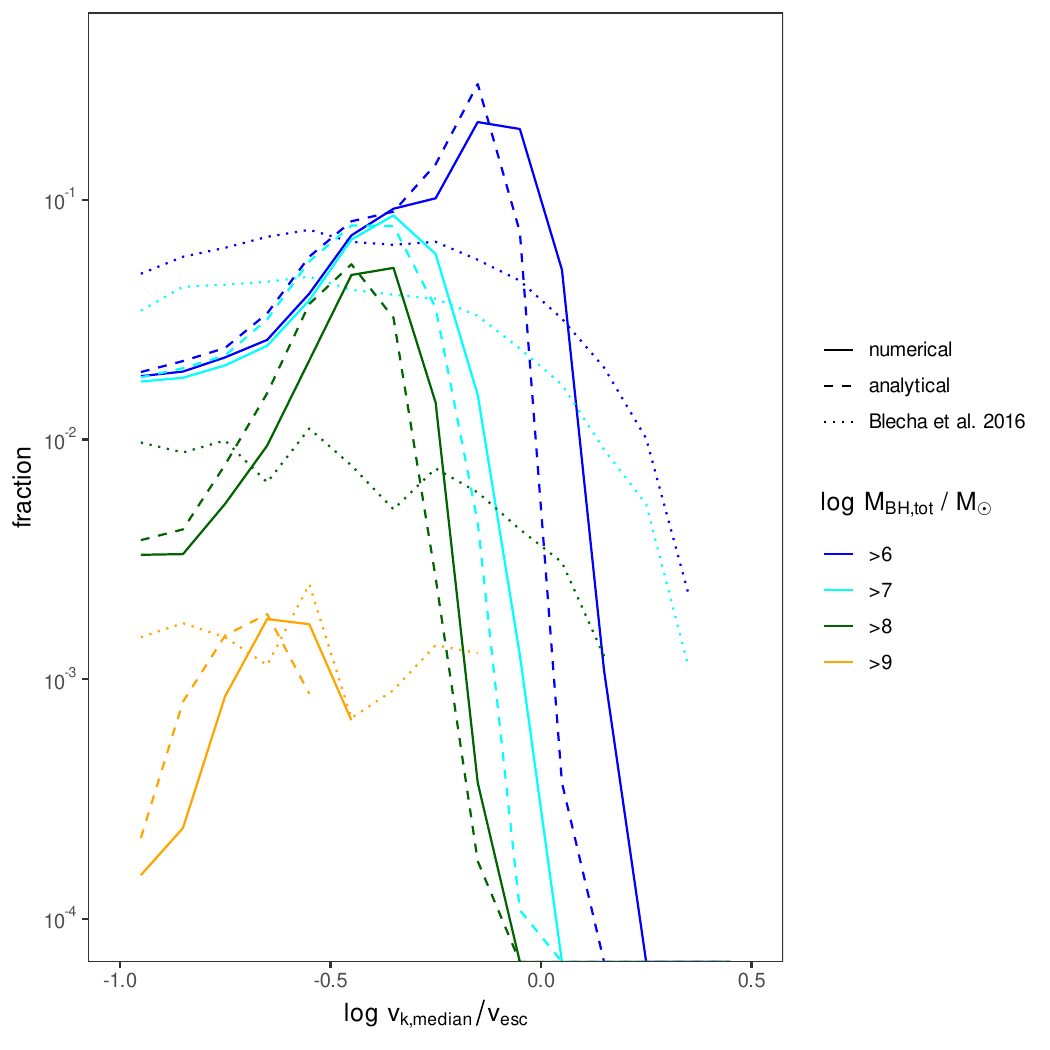}\hfill
\includegraphics[width=.45\textwidth]{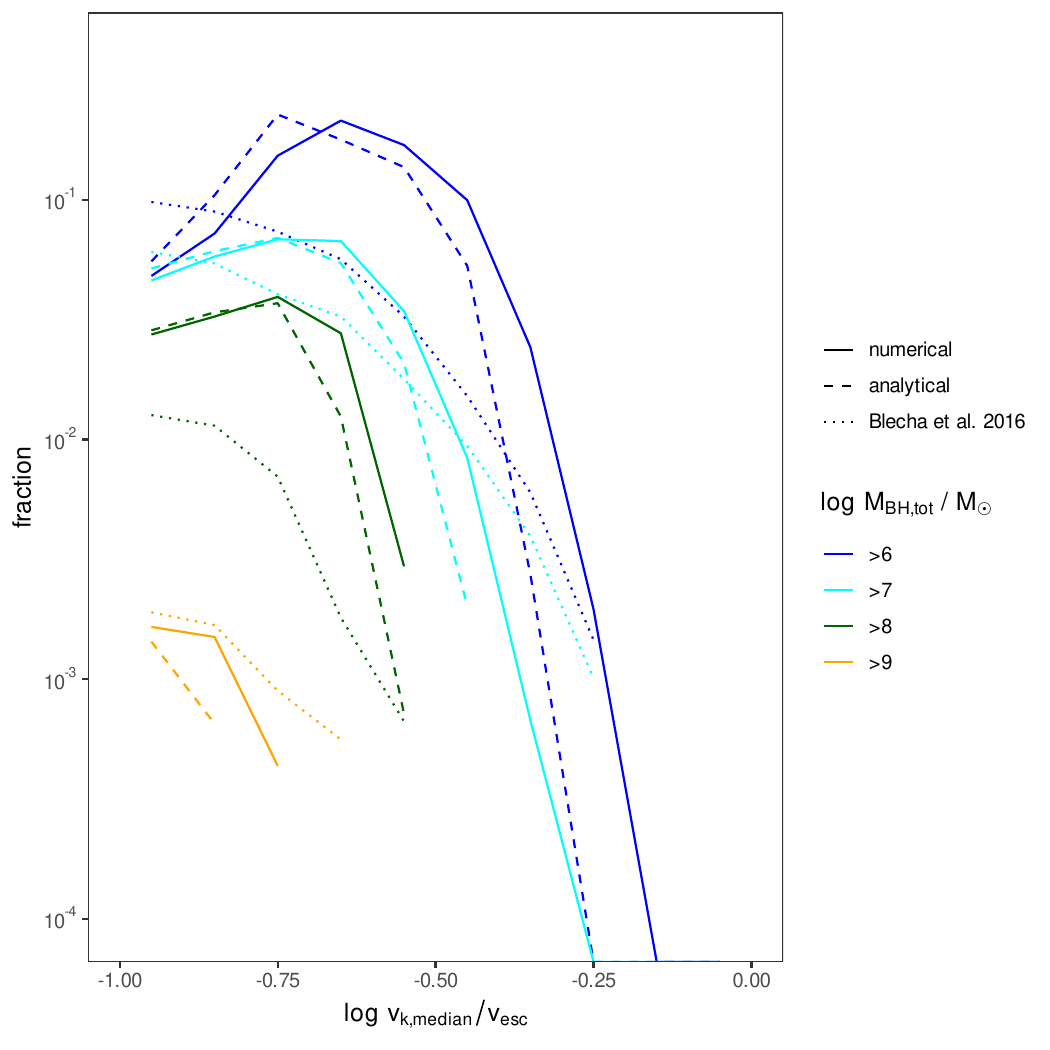}}
    \caption{The distribution of $\log(v_{\textrm{k,median}}/v_{\textrm{esc}})$ ratio for random (left 
    panel) and aligned (right panel) spin distributions. Solid and dashed lines denote
    numerically and analytically calculated escape velocities, respectively. 
    Different colors indicate the total merging BH mass bins.
    The dotted line represents the distribution obtained by Blecha et al. (2016). For kick velocities 
    $v_{\textrm{k,median}}<v_{\textrm{esc}}$ our models predict a greater number of offset AGN. 
}
    \label{offset_frac}
\end{figure*}

Further, we explore the distribution of offset AGN, thus we also take into account recoiling 
BHs with kick velocities lower than the escape velocity, but potentially still
high enough to produce a spatial offset. 
Since \citet{Blecha} provided a similar distribution for the Illustris sample, 
we reproduce their figure 4 and compare our results. 
Fig.~\ref{offset_frac} shows the distribution of median kick velocity for a BH merger pair, 
scaled to the host escape velocity in numerical (solid line) and analytical (dashed line)
models.  
The distribution is separated by total BH mass, indicated by different colors. 
Blue line ($\log(M_{\textrm{BH,tot}}/\Msun)>6$) shows all BH mergers in our sample. 
A greater fraction of escaped BHs is predicted for low-mass BHs, while 
complete BH ejections of most massive BHs are not expected.
However, the plot shows the median kick velocity obtained after 10000 realizations.
Assuming random spin distribution even the most massive BHs occasionally 
get kick velocities large enough for complete ejection from their hosts.
If BH spins are aligned, kick velocities $\log(v_{\textrm{k,median}}/v_{\textrm{esc}})>-0.5$ are possible only 
for the lowest mass BHs. 
Regardless of the total BH mass and the chosen model for BH spin distribution, 
numerical models always predict a greater number of offset AGN. 

Dotted lines on Fig.~\ref{offset_frac} represent the distribution obtained by \citet{Blecha}, assuming 
random and aligned spin distributions.
Our models predict a greater fraction of BHs with kick velocities 
$v_{\textrm{k,median}}<v_{\textrm{esc}}$. Recoiled BHs in 
range $\log v_{\textrm{k, median}}/v_{\textrm{esc}} \in (-0.5,1)$ 
are expected to produce offset AGN that can spend a significant time on bound orbits, thus our models
predict a greater number of offset AGN compared to the findings of \citet{Blecha}. 
However, we note that our results
cannot be compared directly, since our sample contains only major galaxy mergers
from TNG300 simulation,
while \citet{Blecha} shows the distribution of galaxy mergers with all mass ratios from 
lower volume cube of Illustris simulation. In fact,
distributions calculated in this work predict a higher fraction of offset AGN, using both
analytical and numerical estimates for the escape velocity. 
This indicates that the values of kick velocities in our sample are higher. 
Alongside BH spin parameters, the mass ratio of merging BHs plays a key role in
the resulting kick velocity amplitude. Fig.~\ref{vkick_plot} shows that the highest kick velocities can be achieved in mergers
of BHs with comparable masses. In contrast, mergers of low mass ratio binaries result in low kick velocities
and escaped BHs in such systems are expected to be rare.
As shown in Fig.~\ref{q_frac} the distribution of merging BH mass ratios
from our sample has a lower fraction of low mass ratios compared to
\citet{Blecha} sample. By including only major mergers in our sample, 
we have introduced a preference for higher kick velocities.
Again, a lower fraction of most massive BHs in our sample
reflects the imposed upper limit for host galaxy mass of $M_\textrm{DM}<8\times10^{13}\Msun$.

\begin{figure*}
   \resizebox{\hsize}{!}
            {\includegraphics[width=.45\textwidth]{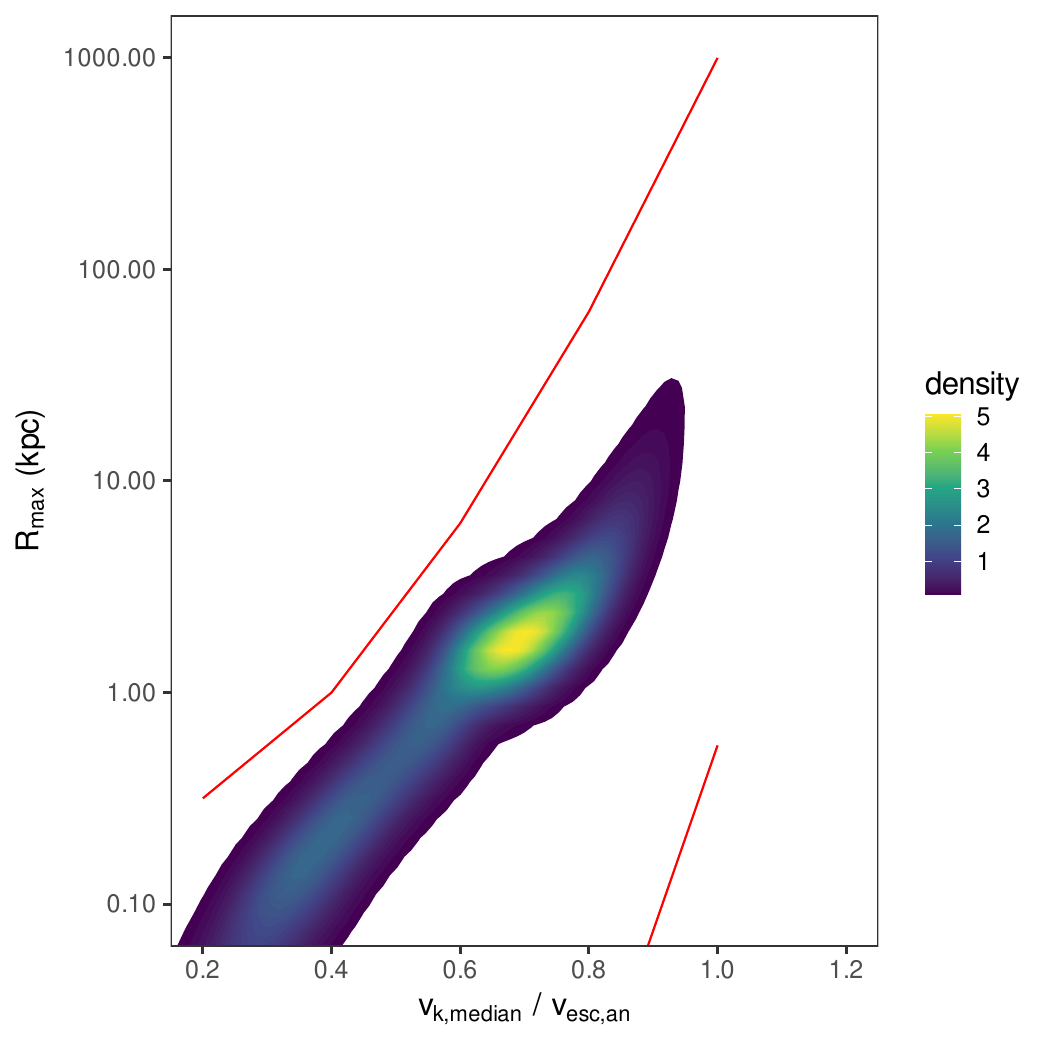}\hfill
\includegraphics[width=.45\textwidth]{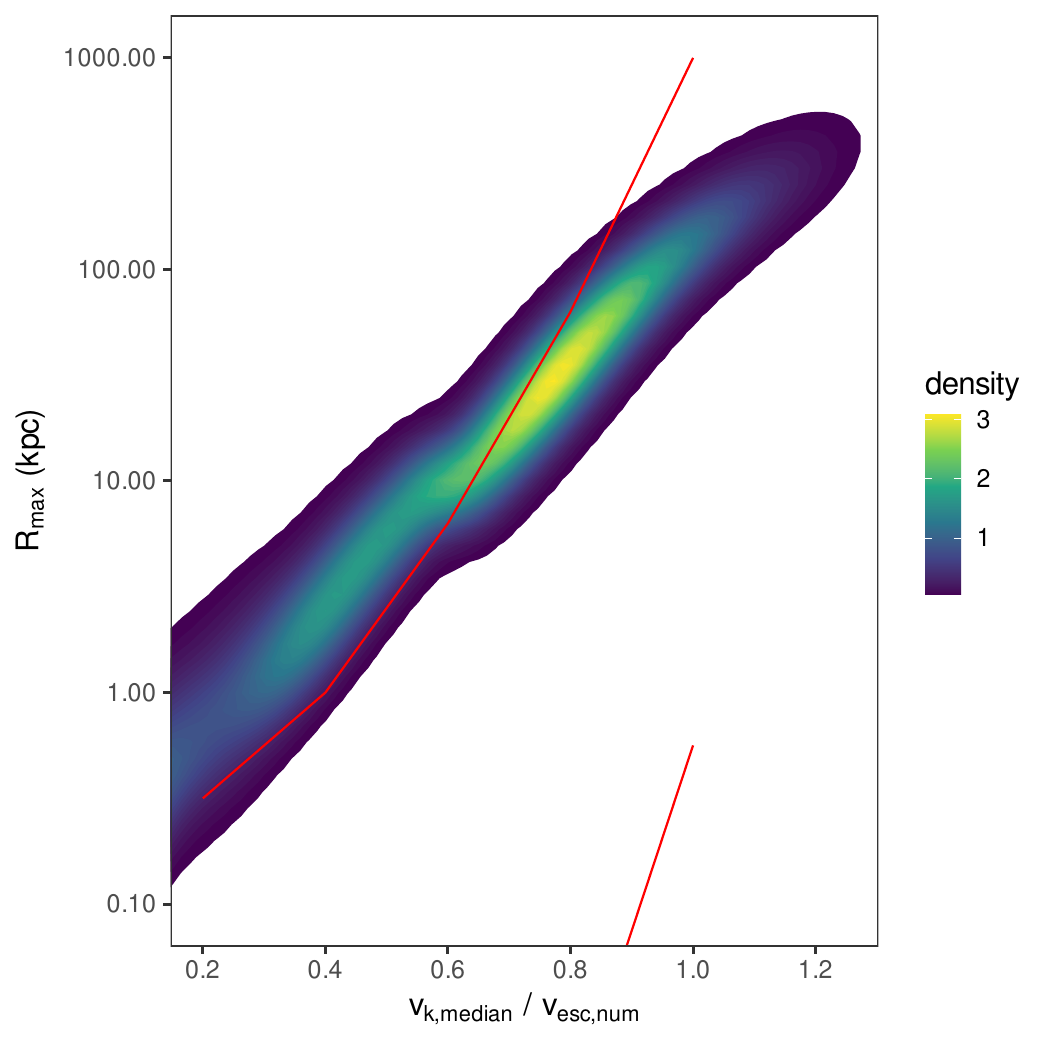}}
    \caption{Maximal separation from a galaxy center reached by a recoiled
    BH as a function of $v_{\textrm{k, median}}/v_{\textrm{esc}}$ ratio for analytical (left) 
and numerical (right) models.
    Red lines represent the upper and lower limits of the distribution obtained by Blecha et al. (2016).
}
    \label{offset_rmax}
\end{figure*}

Next, we explore the maximal distance from a galaxy center reached by recoiled BH, assuming 
the median value of kick velocity. 
Fig.~\ref{offset_rmax} shows the maximal separation from a galaxy
center as a function of $v_{\textrm{k, median}}/v_{\textrm{esc}}$ ratio.
Maximal distances from the host center that BHs could
reach in our analytical models (left panel) are in agreement with 
\citet{Blecha} predictions (red lines).
Recoiled BHs from both TNG300 (this work) and Illustris (\cite{Blecha})
samples will reach comparable separations if their trajectories are 
calculated analytically. This agreement is anticipated as we employ similar analytical models.
In numerical models (right panel), 
due to the violent relaxation process, 
redistribution of mass within the merger remnant galaxies
will allow recoiled BHs to reach greater separations.
BHs on larger galactocentric
distances also spend more time on bound orbits outside of
the galactic center, thus for the given  $v_{\textrm{k, median}}/v_{\textrm{esc}}$ ratio
numerical models predict more spatially offset AGNs.

Table~\ref{rmax_table} shows the total number of recoiled BHs 
at separations $>5$ kpc and $>20$ kpc per redshift bin, assuming numerical versus analytical 
models. 
For a median value of kick velocity majority of recoiled BHs will reach 
separations $>5$ kpc in numerical models, in contrast to analytical models where
$\ge90~\%$ BHs will stay bound to central galaxy regions. 
The total number of offset AGN, as well as the difference between analytical and numerical models, 
increases with redshift, which reflects the hierarchical growth of structures and a
greater number of low-mass galaxies in the early Universe.

\begin{table}[hbt!]
\begin{threeparttable}
\caption{Total number of offset AGN}
\label{rmax_table}
\begin{tabular}{|c|c|c|c|}
\hline
redshift & $R_{\textrm{max}}$ (kpc) & analytical & numerical \\ \hline
all $z$ & \textgreater 5 & 1550 (3\%) & 26722 (58\%) \\ \cline{2-4} 
 & \textgreater 20 & 208 (0.5\%) & 16408 (36\%) \\ \hline
$0 < z {\scriptstyle{\le}} 1$ & \textgreater 5 & 78 (0.5\%) & 6094 (41\%) \\ \cline{2-4} 
 & \textgreater 20 & 9 (0.06\%) & 2695 (18\%) \\ \hline
$1 < z {\scriptstyle{\le}} 2$ & \textgreater 5 & 777 (4\%) & 12246 (68\%) \\ \cline{2-4} 
 & \textgreater 20 & 103 (0.5\%) & 7899 (40\%) \\ \hline
$2< z {\scriptstyle{\le}} 3$ & \textgreater 5 & 595 (7\%) & 6465 (73\%) \\ \cline{2-4} 
 & \textgreater 20 & 77 (0.9\%) & 4568 (51\%) \\ \hline
$3< z {\scriptstyle{\le}} 4$ & \textgreater 5 & 86 (4\%) & 1524 (74\%) \\ \cline{2-4} 
 & \textgreater 20 & 18 (0.9\%) & 1000 (49\%) \\ \hline
$4< z {\scriptstyle{\le}} 5$ & \textgreater 5 & 10 (3\%) & 329 (79\%) \\ \cline{2-4} 
 & \textgreater 20 & 1 (0.2\%) & 202 (49\%) \\ \hline
$z > 5$ & \textgreater 5 & 2  (4\%) & 41 (90\%) \\ \cline{2-4} 
 & \textgreater 20 & 0 & 31 (67\%) \\ \hline
\end{tabular}
\end{threeparttable}
\end{table}

\section{Conclusions}

Static analytical models of merger remnant galaxies have been widely used to study 
recoiling BHs (e.g. \cite{Blecha}). 
However, \citet{smole19} suggested that numerical models represent more realistic 
approach since galaxy mergers can lead to a decrease in galaxy mass
via violent relaxation process that is not depicted in static analytical models.
The authors showed that escape velocities in numerical models 
can be up to 25 $\%$ lower compared to those in analytical models.

Here, we extended the above work and investigated 
the comparative statistics between recoiling BHs in analytical and
numerical models of galaxies extracted from TNG300 simulation.
Our sample consisted of 46031 well-resolved major merger remnant galaxies outside galaxy clusters.
For the given merger remnant galaxy mass, central BH mass, and  mass ratio of progenitor galaxies
we estimated analytical and numerical escape velocities, extrapolating
the escape velocities calculated by \citet{smole19}.
Kick velocities are calculated using the method described by \citet{micic11}.
For each merging BH pair from our sample, BH spins are sampled 
from random or aligned spin distributions, using 10000 realizations.
Comparing kick velocities to the escape velocity from the host galaxy, 
we are able to distinguish between recoiled BHs on bound orbits and escaped 
BHs. 

If BH spins are taken from random spin distribution, 
numerical models predict that 25 $\%$ of merging BHs from our sample will
have escape probability $>40~\%$, compared to $8~\%$ for analytical models.
On the other hand, aligned spin distribution does not yield kick velocities large enough
to produce escaped BHs. 
As redshift increases, the disparity between analytical and numerical models 
becomes more prominent, highlighting the turbulent evolution of galaxies at high redshifts.
Processes in merging galaxies cannot be described using static analytical models alone,
thus numerical models provide a better estimate of the number of escaped BHs.

High escape probabilities predicted by numerical models can 
have a negative influence on BH growth through mergers. 
Major merges of gas reach galaxies can fuel gas into central galaxy regions,
trigger episodes of gas accretion and exponential BH growth.
At redshifts $z>5$, $\sim40~\%$ of major mergers in numerical models will produce a recoiling BH with escape probability 
$>40~\%$. This makes merger-driven BH growth at high redshifts challenging. However, if a
kicked BH remains bound, future mergers are less likely to eject it 
since escape probability decreases at lower redshifts, as galaxies build their mass.
For aligned BH spin model, kick amplitudes are not high enough to 
halt BH growth through mergers, since neither numerical nor analytical models produce
recoiling BHs with escape probability $>2~\%$.

Numerical models also predict a greater number of offset AGN.  
For any given fraction of the escape velocity, recoiled BHs in numerical models are able to reach
greater distances from the host center, increasing the probability of their detection.
Assuming median values of kick velocity numerical models predict that 58 $\%$ of BH mergers in our sample
would produce kicks large enough to displace BHs at separations $>5$ kpc in numerical models,
while analytical models predict only 3 $\%$ of BHs at those separations. 
Recoiled BHs with larger spatial offsets also have longer return times,
which prolongs the period during which they could possibly be observed.

Thus, on cosmological scales, numerically modeled recoiling BHs have a higher escape probability 
and predict a greater number of offset AGN. We conclude that 
numerical models should be favored over analytical ones since BH ejections
take place in non-virialized merger remnants.


\begin{acknowledgement}

\comment{
The python packages \texttt{matplotlib} \citep{Hunter2007}, \texttt{seaborn} \citep{Waskom2021}, \texttt{numpy} \citep{Harris2020}, \texttt{scipy} \citep{Virtanen2020}, \texttt{pandas} \citep{McKinney2010}, and \texttt{pynbody} \citep{pynbody} were all used in parts of this analysis.
}
\end{acknowledgement}

\paragraph{Funding Statement}

This research was supported by the Ministry of Science, Technological Development and Innovation of the Republic of Serbia (MSTDIRS) through contract no. 451-03-47/2023-01/200002 made with Astronomical Observatory of Belgrade.
This research was partially supported by the computing infrastructure acquired in the frame of the project BOWIE (PROMIS 6060916, Science Fund of the Republic of Serbia).

\paragraph{Competing Interests}

None

\paragraph{Data Availability Statement}

The data generated in this study are available from the corresponding author, M.S., upon reasonable request.


\printbibliography

\end{document}